\documentclass[useAMS,usenatbib]{mn2e}

\usepackage{latexsym,amsmath,amstext} 
\usepackage[dvips]{graphicx}          

\usepackage{txfonts}
\usepackage{graphicx}
\usepackage{natbib}
\bibpunct{(}{)}{;}{a}{}{,}

\title[Formation of ``3-D" planetary systems]{Formation of ``3-D" multi-planet systems by dynamical disruption of multiple-resonance configurations}

\author[A.-S. Libert and K. Tsiganis]{A.-S. Libert$^{1,2}$\thanks{E-mail:
anne-sophie.libert@fundp.ac.be (ASL); tsiganis@astro.auth.gr (KT)} and K. Tsiganis$^{2}$\\
$^{1}$Department of Mathematics FUNDP, 8 Rempart de la Vierge, B-5000 Namur, Belgium\\
$^{2}$Department of Physics, University of Thessaloniki, GR-54\ 124, Thessaloniki, Greece}

\begin{document}

\pagerange{\pageref{firstpage}--\pageref{lastpage}} \pubyear{2009}

\maketitle

\label{firstpage}

\begin{abstract}
Assuming that giant planets are formed in thin protoplanetary discs, a ``3-D" system -- i.e. a planetary system composed of two (or more) planets, whose orbital planes have large values of mutual inclination -- can form, provided that the mutual inclination is excited by some dynamical mechanism. Resonant interactions and close planetary encounters are thought to be the primary inclination-excitation mechanisms, resulting in a resonant or non-resonant system, respectively. If by the end of planet formation the system is dynamically ``hot", then a phase of planet-planet scattering can be expected; however, this need not be the case in every system. Here we propose an alternative formation scenario, starting from a system composed of three giant planets in a nearly co-planar configuration. As was recently shown for the case of the solar system, planetary migration in the gas disc (Type II migration) can force the planets to become trapped in a multiply-resonant state (similar to the Laplace resonance in the Galilean satellites). We simulate this process, assuming different values for the planetary masses and mass ratios. We show that, such a triple resonance generally becomes unstable, as the resonance excites the eccentricities of all planets, and  planet-planet scattering sets in. One of the three planets is typically ejected from the system, leaving behind a dynamically ``hot" (but stable) two-planets configuration. The resulting two-planet systems typically have large values of semi-major axes ratio ($\alpha = a_1/a_2 < 0.3$), while the mutual inclination can be as high as 70$^{\circ}$, with a median of $\sim 30^{\circ}$. These values are quite close to the ones recently obtained for the $\upsilon$-Andromedae system. A small fraction of our two-planet systems ($\sim 5\%$) ends up in the stability zone of the Kozai resonance. In a few cases, the triple resonance can remain stable for long times and a ``3-D" system can form by resonant excitation of the orbital inclinations; such a three-planet system could be stable if enough eccentricity damping is exerted on the planets. Finally, in the single-planet resulting systems, which are formed when two planets are ejected from the system, the inclination of the planet's orbital plane with respect to the initial invariant plane -- presumably the plane perpendicular to the star's spin axis -- can be as large as $\sim 40^{\circ}$.
\end{abstract}


\section{Introduction}

The discovery of $\sim 50$ (June 2010) extrasolar multi-planet systems, whose planets have orbital characteristics quite different from the ones in our solar system, has opened new questions about the formation, evolution, and stability of such systems. Because of the lack of spatial resolution of the orbits, as well as a general belief that planetary systems are composed of planets on nearly co-planar orbits, a limited number of studies have addressed the dynamics of ``3-D systems", namely systems composed of two or more planets, whose orbital planes have a significant value of mutual inclination. Such ``3-D" systems can be long-term stable, either following regular secular dynamics or due to the action of some phase-protection mechanism, such as a mean motion resonance (MMR) or a secular Kozai-type resonance (\citealt{Koz62}). 

\begin{table*}   
\centering
\caption{Parameters of the five exosystems analyzed by \citealt{Lib09a}. All the parameters come from \citet{But06}, excepted those of the system HD 155358 from \citet{Coc07}.}
\label{tableparam}
\begin{tabular}{lcccccc}
\hline \hline
&$a$ &  $e$ & $\omega$ (deg) & m sin{i} ($M_{Jup}$) &  $M_{Star}$ ($M_{Sun}$)  \\
\hline
$\upsilon$ And & 0.832 & 0.262 & 245.5 &  1.98 &1.32  \\
$\;\;$ (c-d) & 2.54 & 0.258 & 279 &  3.95 &   \\
\hline
HD 12661 & 0.831 &  0.361 & 296.3 & 2.34 & 1.11    \\
& 2.86 & 0.017 & 38 &  1.83 &   \\
\hline
HD 169830 & 0.817 & 0.310 & 148 & 2.9 & 1.43  \\
& 3.62 & 0.33 & 252 & 4.1 &  \\
\hline
HD 74156 & 0.29 & 0.6360 & 181.5 & 1.8 & 1.21 \\
& 3.35 & 0.583 & 242.4 & 6 &    \\
\hline
HD 155358 & 0.628 & 0.112 & 162 & 0.89 & 0.87  \\
& 1.224 & 0.176 & 279 & 0.504 &  \\
\hline
\end{tabular}
\end{table*}

In \citet{Lib09a}, we studied the possibility that extrasolar two-planet systems, similar to the ones that are observed, can be in a stable Kozai-resonant state. Five known multi-planet systems that are not in MMR were selected as `possible prototypes' ($\upsilon$ Andromedae, HD 12661, HD 169830, HD 74156, HD 155358, see Table \ref{tableparam} for their orbital parameters). Following a parametric numerical study, verified by the analytical secular theory of \citet{Lib07b,Lib08}, we found that four of these systems ($\upsilon$ Andromedae, HD 12661, HD 169830, HD 74156) are consistent with a stable Kozai-resonant state, if their (unknown) mutual inclination is $\sim 45^\circ$. It should be stressed that observational uncertainties and/or incomplete modeling from our part (e.g. absence of general-relativistic precession) are such that one cannot identify Kozai-type motion with the desired certainty. However, our study showed that a good fraction of the detected multi-planetary systems have physical/orbital characteristics compatible with the Kozai state.

The results of \citet{Mc10} for the $\upsilon$-Andromedae system have verified that the orbital planes of planets $c$ and $d$ of this non-resonant system have a mutual inclination of $\sim 30^\circ$. Thus, although the Kozai state is not supported by the observation, the large value of the mutual inclination was confirmed. Thus, the system is near but most likely outside the libration zone of the Kozai resonance. The ratio of the semi-major axes of the two planets is such that orbital intersections are avoided ($\alpha = a_c/a_d \approx 0.32$) and so the secular dynamics are regular.

Resonant systems of two planets can also be ``3-D", since the orbital inclinations can grow under the action of MMRs. This was first shown by \citet{Tho03}, who studied the evolution of planets trapped in a 2:1 MMR, under the effects of gas-driven (Type II) migration, inside the protoplanetary gas disc. \citet{Tho03} found that, once the eccentricities are high enough, the system that is captured in 2:1 MMR also enters an ``inclination-type" resonance, which induces rapid growth in the inclinations of both planets. This work was extended by \citet{Lib09b}, where it was shown that capture in higher-order resonances (such as the 5:2, 3:1, 4:1 and 5:1) are possible for a wide range of migration and eccentricity damping rates (i.e. $\dot{a}$ and $\dot{e}$). Moreover, it was shown that these MMRs are also able to excite the inclinations to high values, provided that eccentricity damping is not very strong, so that at least one of the planetary orbits has an eccentricity higher than $e=0.4$. The conclusion of that work was that the inclination-excitation mechanism can be quite common in resonant systems and thus a large number of them may in fact represent cases of ``3-D" systems. We stress here the fact that resonant trapping due to Type II migration is considered as the main formation mechanism for resonant multi-planet systems. 

Dynamical instabilities of systems with giant planets have been proposed to explain the orbital properties of extrasolar systems, in particular the highly eccentric orbits seen in many systems. \citet{For05} showed that the current orbital configuration of the giant planets in the $\upsilon$-Andromedae system probably resulted from planet-planet scattering with an additional planet, now lost from the system. A similar mechanism, accompanied by tidal circularization of the resulting highly eccentric orbits (\citealt{For08}), is commonly invoked to explain the formation of ``hot Jupiters" (i.e.\ massive planets on nearly circular orbits and with semi-major axes smaller than 0.03~AU). \citet{Nag08} also showed that a combination of planet-planet scattering, tidal circularization and the Kozai mechanism in systems with three Jupiter-mass planets may explain the observed frequencies of hot Jupiters (see also the work of \citealt{Fab07}).

The formation of non-resonant ``3-D" systems (like the ones in Table \ref{tableparam}) -- more precisely, the increase of the mutual inclination of the orbital planes of the planets -- is also thought to originate from this violent dynamical mechanism, namely planet-planet scattering. This is in fact the same mechanism that is generally invoked to explain the eccentricity distribution of extrasolar planets. Most of the previous studies on multi-planet scattering were focussed on gas-free systems composed of two planets, aiming at reproducing the observed eccentricity distribution of extrasolar planets (see for instance \citealt{For01}, \citealt{For08}). \citet{Mar02} explored the stability and final orbital properties of three-planet systems, showing that the most common outcome of gravitational scattering by close encounters is hyperbolic ejection of one planet, the two ``survivors" having significant values of eccentricity and mutual inclination. \citet{Cha07} and \citet{Jur08} extended this work, showing that planet-planet scattering can reproduce quite well the observed eccentricity distribution. 

Moreover, \citet{Cha07} have reported results on series of  planet-planet scattering simulations, where high mutual inclinations (as high as $40^\circ-60^\circ$) between the orbits of the surviving planets were observed. In such simulations, the number of planets initially considered is typically higher than the number of surviving planets, as the price for reaching a stable final configuration is typically the ejection of one (or more) planet from the system. We note also that \citet{Jur08} studied the systems with more than three planets, finding that the number of surviving planets in their simulations was typically 2 or 3, suggesting that extrasolar planetary systems are unlikely to harbor more than three giant planets. At present, multi-planet scattering seems to be the most promising scenario for forming non-resonant ``3-D" systems with $I_{mut}>30^{\circ}$. 

The starting point of planet-planet scattering simulations usually is that the systems are relatively compact when the gas dissipates, such that they can become unstable on relatively short time scales. Only few works investigated the combined action of disk torques and planet-planet scattering (e.g. \citealt{Ada03, Moo05}, for preliminary studies of two-planet systems). \citet{Tho08} studied the formation of giant planets, assuming different values of the disc parameters (the planet-disc interactions were modeled using a $N$-body code, combined with a 1-D disc model). In that work, it was shown that gas-driven migration tends to produce crowded systems, in which eccentricity excitation due to either resonances or planet-planet scattering occurred. However, \citet{Mat10} suggested that, according to their simulations (combining $N$-body dynamics with hydrodynamic disc evolution), planet-planet scattering conditions are difficult to achieve. In fact, the existence of a large fraction of resonant systems suggests that unstable systems (i.e. systems in which planet-planet scattering sets in immediately after the gas dissipates) is not a unique result of the formation process. Of course, as is clear from this discussion, it is not easy to determine a priori what fraction of systems (and under which conditions) will reach a stable or a 
marginally unstable configuration, leading to planet-planet scattering.

A possible ``intermediate" evolutionary path has been noted recently, in studies devoted to the origin of the solar system. \citet{Morbicrida07} studied the migration of the Jupiter-Saturn pair (also pairs of heavier planets) in gas discs with different characteristics. They showed that, at least for systems where the outer planet has a smaller mass than the inner one, trapping into a low-order MMR is accompanied by a significant drop (or even sign reversal) of the migration rate. Equal-mass planets (Jupiter mass) are also captured in resonance, but the migration direction does not change. Thus, the Jupiter-Saturn was likely trapped in a 3:2 resonance during the gas phase and then stopped migrating. The results of this study were used by \citet{Mor07}, who demonstrated that the remaining outer planets of the solar system (Uranus and Neptune) could have followed a similar evolution, each one being trapped in a low-order MMR (3:2, 4:3 or 5:4) with the immediately preceding planet. Thus, at least in the case of the solar system, gas-driven migration can force the planets to enter into a {\it multiple resonance}, an analogue of the Laplace resonance in the Galilean moons. As noted in \citet{Mor07}, a multiple planetary resonance is a delicate dynamical configuration; not all resonant ratios can be reached by all planetary masses (and mass ratios) and not all resonances are long-term stable. As shown in \citet{Mor07} resonant interaction can increase the planetary eccentricities, such that planet-planet scattering becomes possible. Then, the (chaotic) dynamics will determine the orbital configuration in the final, likely ``3-D", system. Moreover, even if the multiple resonance remains stable, inclination excitation may occur (as in \citealt{Lib09b}), again resulting into a ``3-D" (resonant) system.

In this paper, we wish to explore this mechanism, in the context of extrasolar systems; in particular, the formation of ``3-D" multi-planet systems. The main point of interest in this mechanism is that it may apply to multi-planet systems that are not formed in a dynamically unstable configuration. In particular, if planets form on quasi-circular orbits that are sufficiently separated, the system can be stable on a $\sim 10$~My time scale, which may be long enough for differential migration in the gas disc to ``lock" them in multiple resonance\footnote{This can certainly be true for some range of masses and mass ratios.}. On the contrary, it assumes that migration can lead to the formation of a multiple resonance among the planets, which can (and most likely will) become unstable, due to eccentricity excitation induced by the resonance itself. Hence, the scope of the present paper is to examine the evolution of three-planet systems, trapped in a triply-resonant configuration, for planet masses and mass ratios similar to the ones observed in non-resonant two-planet systems (see Table 1). We stress again that the conditions under which a multiply-resonant system with three (or more) giant planets -- possibly much heavier than Jupiter -- can form have to be investigated thoroughly in the future. The aim of our work is to describe the general behavior and orbital distribution of two-planet systems, resulting from the violent dissolution of a triply-resonant configuration, induced by gas-driven migration. A more detailed analysis, both in terms of modeling and statistics, is reserved for future work. Here, we present a first approach to this problem, by simulating the establishment of a 1:2:4 resonance and performing a first statistical study of the orbital characteristics of the resulting systems. Let us note that the choice of this resonance seems in agreement with the observational data : at least the exosystems HD 82943 (\citealt{Beau08}), HR8799 (\citealt{Mar08}), and Gliese 876 (\citealt{Riv10}) may exhibit the Laplace resonance. In the following section, we describe the set-up of our numerical experiments. The results are presented in Section 3 and our conclusions are given in Section 4.

\begin{table*}   
\caption{Percentage of surviving two-planet systems in our simulations of three planets in multiple resonance. For each mass configuration, the percentage of surviving two-planet systems with a mutual inclination of the orbital planes higher than $40^\circ$ is also shown.}
\centering
\begin{tabular}{|c|c|c|c|c|c|c|}
\hline \hline
$m_{2}$ & \multicolumn{2}{|c|}{$m_{3}=2\; m_{2}$} & \multicolumn{2}{|c|}{$m_{3}=m_{2}$} & \multicolumn{2}{|c|}{$m_{3}=1/2\; m_{2}$}\\
\hline
& percentage & percentage & percentage & percentage & percentage & percentage\\
& of surviving & of surviving & of surviving & of surviving & of surviving & of surviving \\
& 2-planet systems & 2-planet systems & 2-planet systems & 2-planet systems & 2-planet systems & 2-planet systems \\
&            & with $I_{mut}\ge 40^\circ$ &            & with $I_{mut}\ge 40^\circ$ &            & with $I_{mut}\ge 40^\circ$ \\
\hline
$ 3$ & \multicolumn{2}{|c|}{no multiple resonance} &  $80\%$ & $30\%$ & $30\%$ & $0\%$  \\
\hline
$1.5$ & $30\%$ & $0\%$& $60\%$& $40\%$ & $60\%$ & $20\%$\\
\hline
$1.15$ & $20\%$ & $0\%$ & $70\% $ & $30\%$ & $50\%$ & $10\%$\\

\hline
\end{tabular}
\label{tablescenario}
\end{table*}

\section{The Model -- Numerical Set-Up}

Our initial (fictitious) systems consist of three giant planets that are forced to migrate such that a triple resonance can be reached. Then, the evolution of each system is followed, until a final, stable, system is formed. The masses and mass ratios of the planets were chosen such that they are close to the values of the non-resonant two-planet systems shown in Table 1. We remind the reader that \citet{Lib09a} found these systems to be stable for high values of mutual inclination. Thus, assuming the mass  of the star to be $M_*=1~M_{\odot}$, we set $m_{1}=1.5$ Jupiter masses. The mass of the second planet is set to $m_{2}=1.15,~1.5$ or $3$ Jupiter masses, while $m_{3}=m_{2}/2, m_{2}$ or $2~m_{2}$. Thus, in total we consider nine different mass configurations. Our main focus is on the semi-major axis and inclination distribution of those final systems that host two planets.  

The starting point in our simulations is the establishment of the triple resonance. Following the lines of \citet{Mor07}, we assume a two-step formation of the multiple resonance. First, we simulate the resonant capture of the first two planets (in order of increasing distance from the star), $m_1$ and $m_2$. For reasons of simplicity we will only study  capture in the 1:2 MMR, in this paper. The initial values of the semi-major axis of these two planets were $a_1=1$ and $a_2=1.9$. Our system of units is such that $G=1$ and $M_{\odot}=1$; thus, the period of a planet with $a=1$~(AU) is 
$T_1=2\pi$. Both initial eccentricities and inclinations are quasi-null ($e=0.001$, $i=0.01^\circ$). All simulations are performed using the numerical integrator described in \citealt{Lib09b}, in which a suitable Stokes-type drag force is added in the equations of motion of the $N$-body problem (following \citealt{Beau05}), to simulate the exponential drift in semi-major axis (migration) and eccentricity (damping). The migration and eccentricity damping rates are set to $1/\tau_a = 2.5\times 10^{-6}$ and $1/\tau_e = 5\times 10^{-6}$ time units, respectively; these values were taken from \citealt{Lib09b}. For this set of parameters, the planets are captured in the resonance after $\sim 10^5~T_1$, when $a_1\approx 0.9$ and $a_2\approx 1.5$. 

After $m_1$ and $m_2$ reach a stable 2:1 configuration, $m_3$ is introduced in the simulation (with $a_3=3$, $e_3=0.001$ and $i_3=0.01^\circ$) and forced to migrate into the 1:2 MMR with $m_2$, using the same recipe as above; hence the triple resonance 1:2:4 is reached. In this second step in each simulation, the drag terms are switched off in the equations of motion of $m_1$ and $m_2$ and only $m_3$ suffers migration; this is done to prevent $m_1$ and $m_2$ from migrating too close to the star, while $m_3$ slowly approaches the resonance. Not applying migration to $m_1$ and $m_2$ also mimics the behavior shown in \citealt{Morbicrida07}, where it is shown that the migration of two giant planets can significantly slow down once they become trapped in resonance, as a result of severe gas depletion in the interplanetary region (formation of a ``common gap"). The triple resonance is reached (but not always established, see next section) after another $\sim 10^5~T_1$. The evolution of the three-planet system is then followed for another $5\times 10^5~T_1$. In most cases, the system becomes unstable during this period, as the resonance pumps the eccentricities of the planets. In order to improve our statistics on the orbits of the final systems, we increased our sample by cloning each unstable run 10 times, taking the orbital parameters of the planets a bit before the instability time and adding small, random, deviations in the velocities of the planets. Two configurations did not go unstable during the integration time ($\sim 6\times 10^5~T_1$). For these cases, we cloned the final conditions of the planets and continued the integration until the cloned systems became unstable. Thus, our sample consists of 90 runs in total. During the instability phase (planet-planet scattering) the drag terms are switched off and the evolution of the three-planets system is followed by a normal $N$-body simulation. As expected, hyperbolic ejection of one planet is the typical outcome. We then focus our attention on the orbital distribution, in particular the mutual inclination, of the resulting two-planet systems.

Before we present our results some comments about our modeling procedure are in order. We fully realize that our modeling is far from complete. As mentioned earlier, it is not clear (and needs to be verified by hydrodynamical simulations) whether all possible three-planet configurations can lead to the formation of a triple resonance. Here, we only study the 1:2:4 configuration, as the ``prototype" of a triple resonance. However, the masses of the planets and the parameters of the disc will determine if and which triple resonance can be established. Concerning also the time-line of evolution, there are two more things to consider. A more gradual decay of the gas disc should probably be considered, rather than an abrupt switching-off of the drag terms in the equations of motion of the planets, during the instability phase. However, we do not consider this to be crucial, as our systems become unstable before this abrupt ``decay" of the disc. On the other hand, no eccentricity or inclination damping is applied during the planet-planet scattering phase, which might be required for systems that reach this phase early enough, i.e.\ before the disc actually dissipates. We note, however, that it is not easy to decide ``how much" damping one should consider for planetary orbits with inclinations much larger than the width of the disc (i.e. planets moving mostly outside the gas disc); at least we are not aware of hydrodynamical simulations addressing this issue. Given the above, it is likely that our quantitative results are not very accurate, but we are confident that the qualitative behavior is correctly reproduced.

\section{Results}

\begin{figure*}
\hspace{-1cm}\rotatebox{0}{\includegraphics[height=12cm]{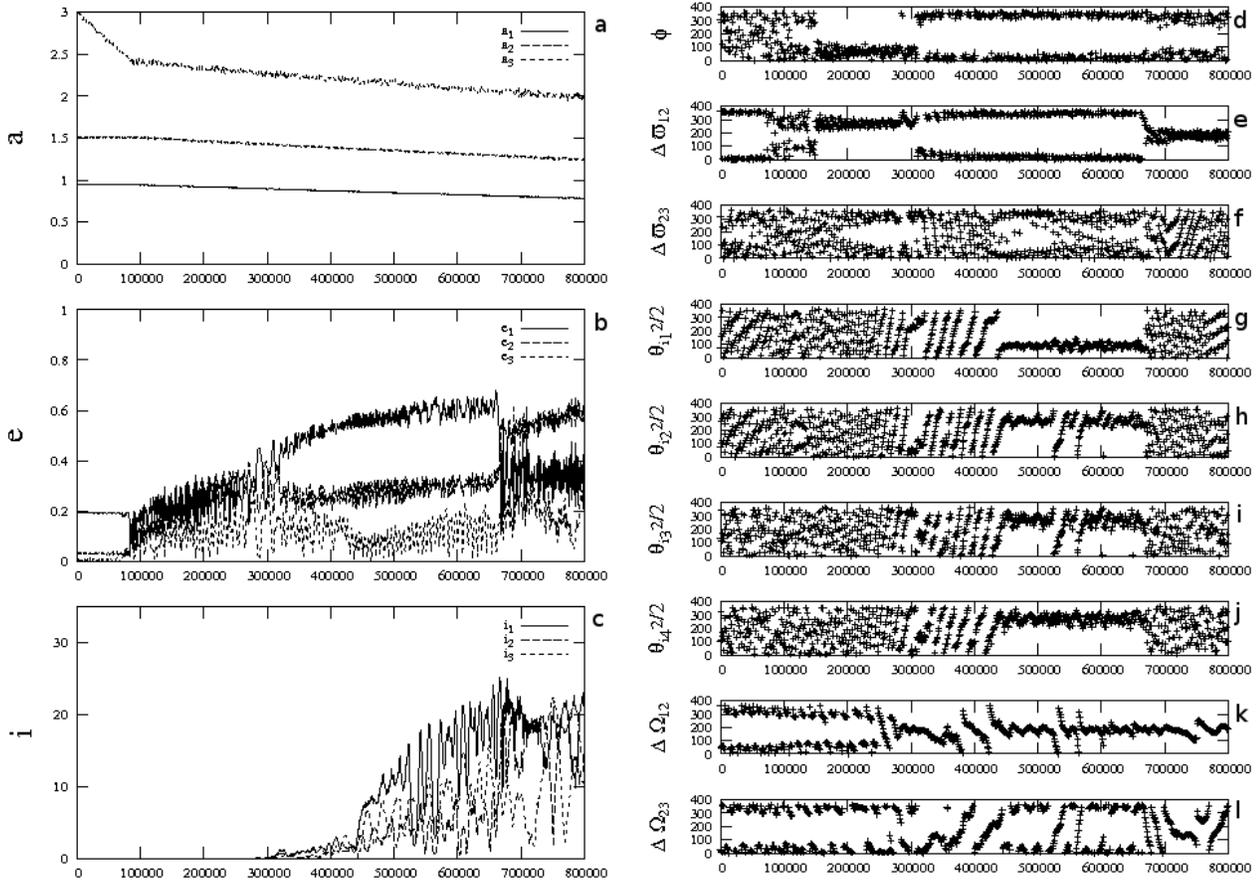}}
\caption{Three-planet system in a 1:2:4 resonant configuration. The planetary masses are $m_1=1.5, m_2=1.5, m_3=0.75 M_{Jup}$. As in the \citet{Tho03} mechanism, we observe an increase of the orbital inclinations for eccentricities greater than 0.2. This mechanism is described in more detail in the text. }
\label{thommes_fig}
\end{figure*} 

Table \ref{tablescenario} summarizes the results of our simulations. For $m_{2}=3M_{Jup}=m_{3}/2$, the systems became unstable before reaching a multi-resonant state; thus, the 1:2:4 resonance cannot be established for this mass 
configuration. This is because the planets are so heavy that, as $m_3$ approached the 1:2 MMR with $m_2$, the orbits become chaotic and planet-planet scattering dissolves the system. This was also observed in some of the simulations of \citet{Lib09b}. In the following we will not consider these systems any more. 
 
For all the remaining mass configurations, a multiple resonance was established, according to the scheme described above. As the eccentricities of the planets grew, most of the systems evolved towards an unstable configuration, reaching eventually the required conditions for planet-planet scattering. The typical outcome was the ejection of one of the planets. The percentage of surviving two-planet systems is given in Table \ref{tablescenario} for each mass configuration. Overall $\sim 50\%$ of all simulations resulted in a two-planet system. As expected, more massive planets tend to survive, while less massive planets tend to be ejected. Also, one of the two ``survivors" moves to a smaller orbital radius (closer to the star) while the other is left on a distant orbit; less massive planets are typically scattered away from their initial orbits, while the more massive ones remain near their initial locations.  

Hereafter we describe the orbital characteristics of the resulting systems. We characterise as {\it stable} a system whose evolution remains bounded and quasi-periodic for an additional numerical integration of $10^6~T_1$ (or $2 \times 10^7~T_1$ in case of 3-D two-planet surviving systems). In some cases, the secular instability of two-planet and three-planet systems might build up over much longer time scale. However, we believe that it would only reduce the number of surviving systems but not modify the general trend of the statistics on their orbital characteristics. In the following, we mainly focus on two-planet systems. We note however that one-planet as well as three-planet stable configurations were found. Finally, a set of 12 runs resulted in two-planet systems with one of the planets spending a long time on a highly-eccentric orbit ($e\sim 0.9$). Had we extended the integration these systems would eventually dissolve, as the planets were still strongly interacting. Since we are focused on stable two-planet systems we decided to discard these runs. Moreover, these runs would not provide reliable results, concerning the surviving one-planet systems, as our dynamical model is not complete enough to account for such long-lived, highly-eccentric orbits.

Before presenting the results on two-planet systems a short note on the dynamics of 1-planet and 3-planet configurations should be given.

\subsection{Three-planet systems}

In the case of mild (or no) eccentricity damping, the triple resonant configuration typically becomes unstable, since the eccentricities of the planets keep increasing as migration continues. However, if the disc dissipates before the system dissolves, a stable configuration can be reached. Nine of our runs remain in a stable 1:2:4 resonant configuration for the whole integration time span (and for an additional numerical integration of $10^6~T_1$). What is interesting in these simulations is that the inclinations of the three planets increase to relatively high values. As shown in Fig. \ref{thommes_fig}, the inclinations start growing after the eccentricities of the planets become greater than $e\sim 0.2$ and reach values of $\sim 25^{\circ}$ when one of the eccentricities exceeds $\sim 0.5-0.6$. This behavior is quite similar to the one described by \citet{Tho03}, in the case of two planets trapped in a 1:2 MMR. Apparently, a similar mechanism is also active in triply-resonant configurations, such as the one studied here. We note that the percentage of systems following this type of evolution (i.e.\ ``3-D" resonant three-planet systems) could be larger, had we assumed stronger eccentricity damping.

To study the inclination excitation mechanism of three-planet systems in more detail, Fig. \ref{thommes_fig} describes the evolution of the critical angles of this type of evolution. The two inner planets are initially captured in a 1:2 MMR, characterized by the libration of both resonant angles $\theta_1=\lambda_1-2\lambda_2+\varpi_1$ and $\theta_2=\lambda_1-2\lambda_2+\varpi_2$ around $0^\circ$. It also means that the difference of the longitudes of the pericenter $\Delta \varpi_{12} = \varpi_1 - \varpi_2 = (\theta_1-\theta_2)/2$ oscillates around $0^\circ$, i.e. the planets are in apsidal alignement (see panel e). The outer planet migrates until the capture in a 2/1 MMR with the second body at about $0.9\times 10^5$ yr. Only one of the two resonant angles, $\theta_3=\lambda_2-2\lambda_3+\varpi_2$ and $\theta_4=\lambda_2-2\lambda_3+\varpi_3$, is in libration, which explains that $\Delta \varpi_{23} = \varpi_2 - \varpi_3 = (\theta_3-\theta_4)/2$ does not oscillate in a first time (panel f). Thus the system is captured in a Laplace-type resonance, whose critical angle is $\phi=\lambda_1-3\lambda_2+2\lambda_3$ (panel d). 

As the three planets continue to migrate while in resonance, their eccentricities increase (panel b). When their values are high enough, the system enters an inclination-type resonance: the angles $\theta_{i_1^2}=2\lambda_1-4\lambda_2+2\Omega_1$, $\theta_{i_2^2}=2\lambda_1-4\lambda_2+2\Omega_2$, $\theta_{i_3^2}=2\lambda_2-4\lambda_3+2\Omega_2$ and $\theta_{i_4^2}=2\lambda_2-4\lambda_3+2\Omega_3$ start to librate around $180^\circ$ at $4.5\times 10^5$ yr (panels g,h,i,j). A rapid growth of the inclinations of the three planets is observed, as well as the librations of the relative longitudes of the nodes, $\Delta \Omega_{12}=\Omega_1-\Omega_2=(\theta_{i_1^2}-\theta_{i_2^2})/4$ and $\Delta \Omega_{23}=\Omega_2-\Omega_3=(\theta_{i_3^2}-\theta_{i_4^2})/4$ (panels k,l).

At $6.75\times10^5$ yr, the system leaves the inclination-type resonance, as $\theta_{i_1^2}$, $\theta_{i_1^2}$, $\theta_{i_1^2}$ and $\theta_{i_1^2}$ stop librating. Note that the relative longitudes of the nodes $\Omega_{12}$ and $\Omega_{23}$ still librate. The resonant angle $\theta_1$ switches his libration center (so does $\Delta \varpi_{12}$, see panel e) but the system remains stable due to the large values of mutual inclinations between the planets. The system still evolves in the Laplace-type resonance until the end of the simulation reproduced in Fig. \ref{thommes_fig}. We expect this configuration to become unstable as the migration continues. Of course, the dissipation of the gas disc, sometimes after resonance capture, could prevent the planets from this unstable phase and form a stable non-coplanar three-planet system.

\subsection{One-planet systems}

\begin{figure}
\hspace{-1cm}\rotatebox{270}{\includegraphics[height=18.cm]{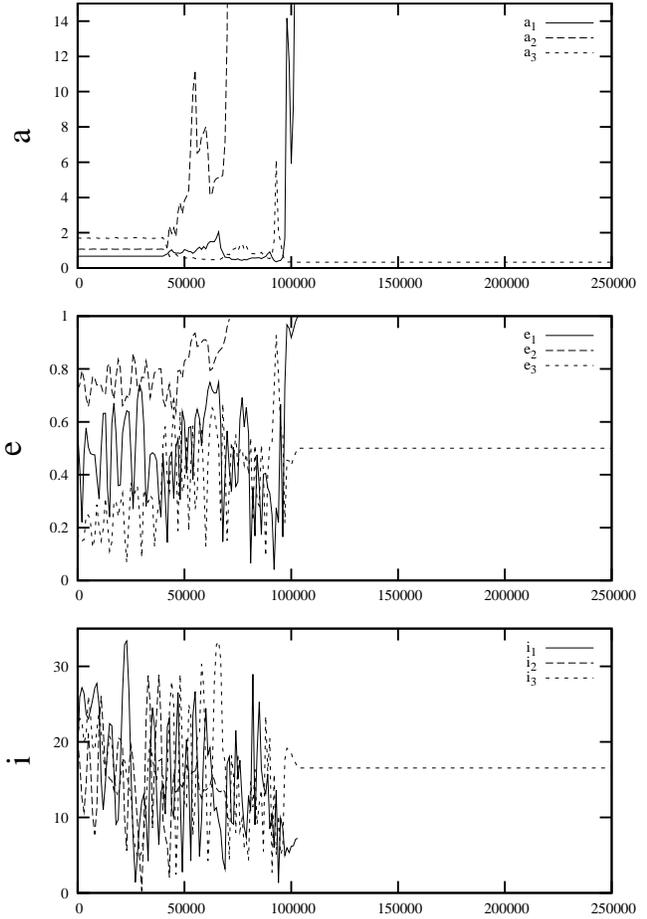}}
\caption{One-planet system obtained by the successive ejections of planet $m_2$ at $71,000~T_1$ and planet $m_1$ at $103,000~T_1$ of the initial three-planet configuration. The planetary masses are $m_1=1.5, m_2=1.5, m_3=1.5 M_{Jup}$. The inclination of the planet's orbital plane with respect to the initial invariant plane -- presumably the plane perpendicular to the star's spin axis -- reaches $\sim 15^{\circ}$ in this simulation.}
\label{onepl_fig}
\end{figure}

About $25\%$ of our simulations (19 runs) result in one-planet systems. Figure \ref{onepl_fig} shows such a behavior: after the ejection of a first planet, the remaining pair is still dynamically unstable and one of the remaining planets is eventually ejected from the system. In almost all cases it is the heavier planet that survives. Although we do not have enough runs to make a proper statistical analysis, an interesting result is found, concerning the orbital inclination of the remaining planet. In 5 cases the final inclination is higher than $15^{\circ}$ with respect to the initial invariant plane (the common orbital plane of the three planets and the presumed gas disc), which can be considered as the plane perpendicular to the spin axis of the host star. The highest value of inclination recorded is $38^{\circ}$.

\subsection{Two-planet systems}

\begin{figure}
\hspace{-1cm}\rotatebox{270}{\includegraphics[height=18.cm]{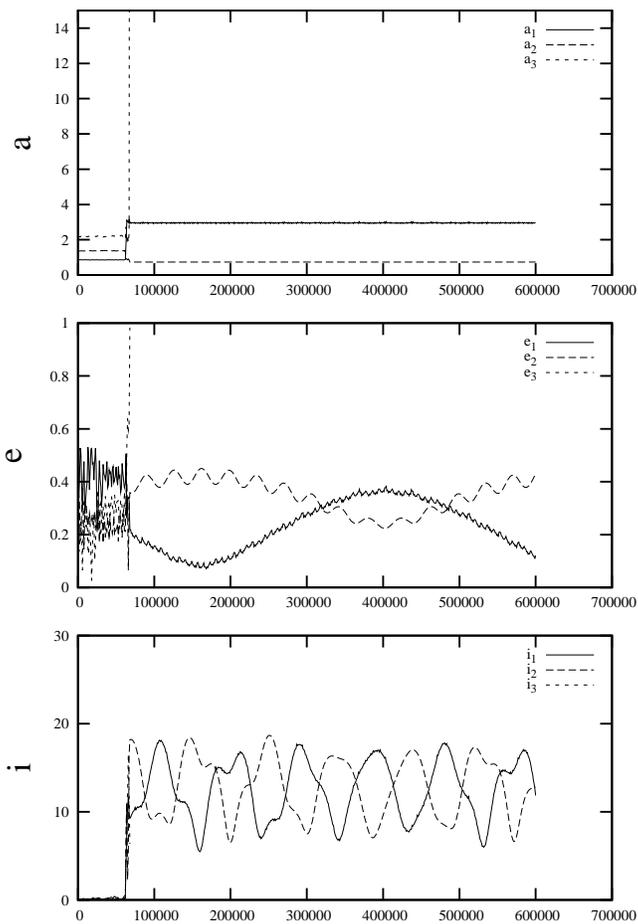}}
\caption{Two-planet system formed by dynamical disruption of a triple resonant configuration on initially nearly co-planar orbits. The planetary masses are $m_1=1.5, m_2=3, m_3=1.5 M_{Jup}$. After the hyperbolic ejection of the less massive outer planet, the remaining system is stable, with a large orbital separation between the two planets, large eccentricities and a large mutual inclination.}
\label{scattering_fig}
\end{figure}

In $\sim 50\%$ of our runs (i.e. 40 runs), the disruption of the triple resonance leads to planet-planet scattering that results in a two-planet system. The final systems are characterised by a large orbital separation ($\alpha = a_1/a_2$ small) between the two planets, large eccentricities and a large mutual inclination. An example is shown in Fig.\ref{scattering_fig}. About $30\%$ of the systems have $I_{mut}>40^{\circ}$, while the median is $30^{\circ}$; this is actually the value found by \citet{Mc10} for the $\upsilon$-And system. This result confirms our suggestion that ``3-D" systems are naturally formed by the dynamical dissolution of a three-planet, multi-resonant configuration, on initially nearly co-planar orbits.

\begin{figure}
\rotatebox{270}{\includegraphics[height=10.cm]{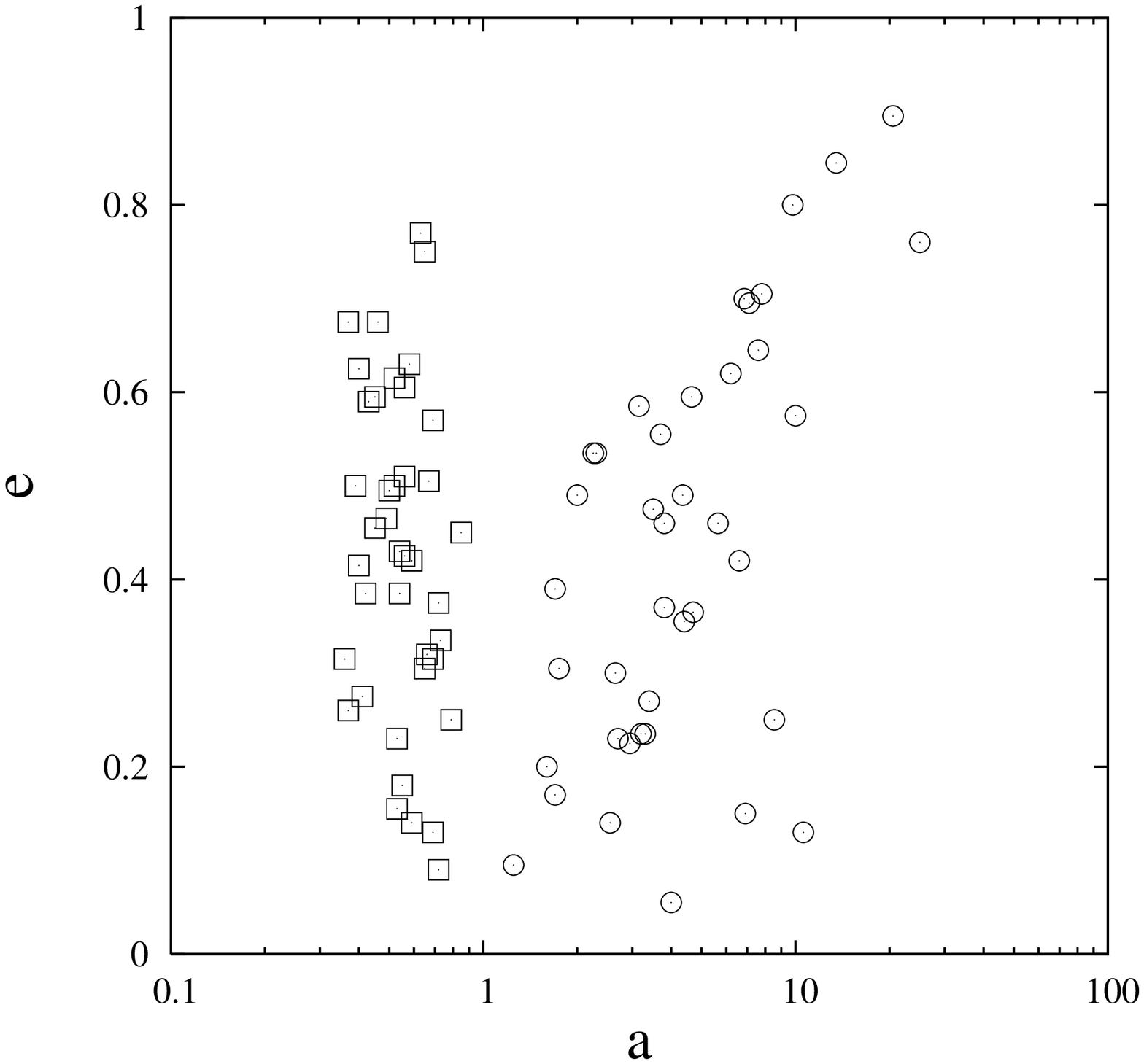}}
\rotatebox{270}{\includegraphics[height=10.cm]{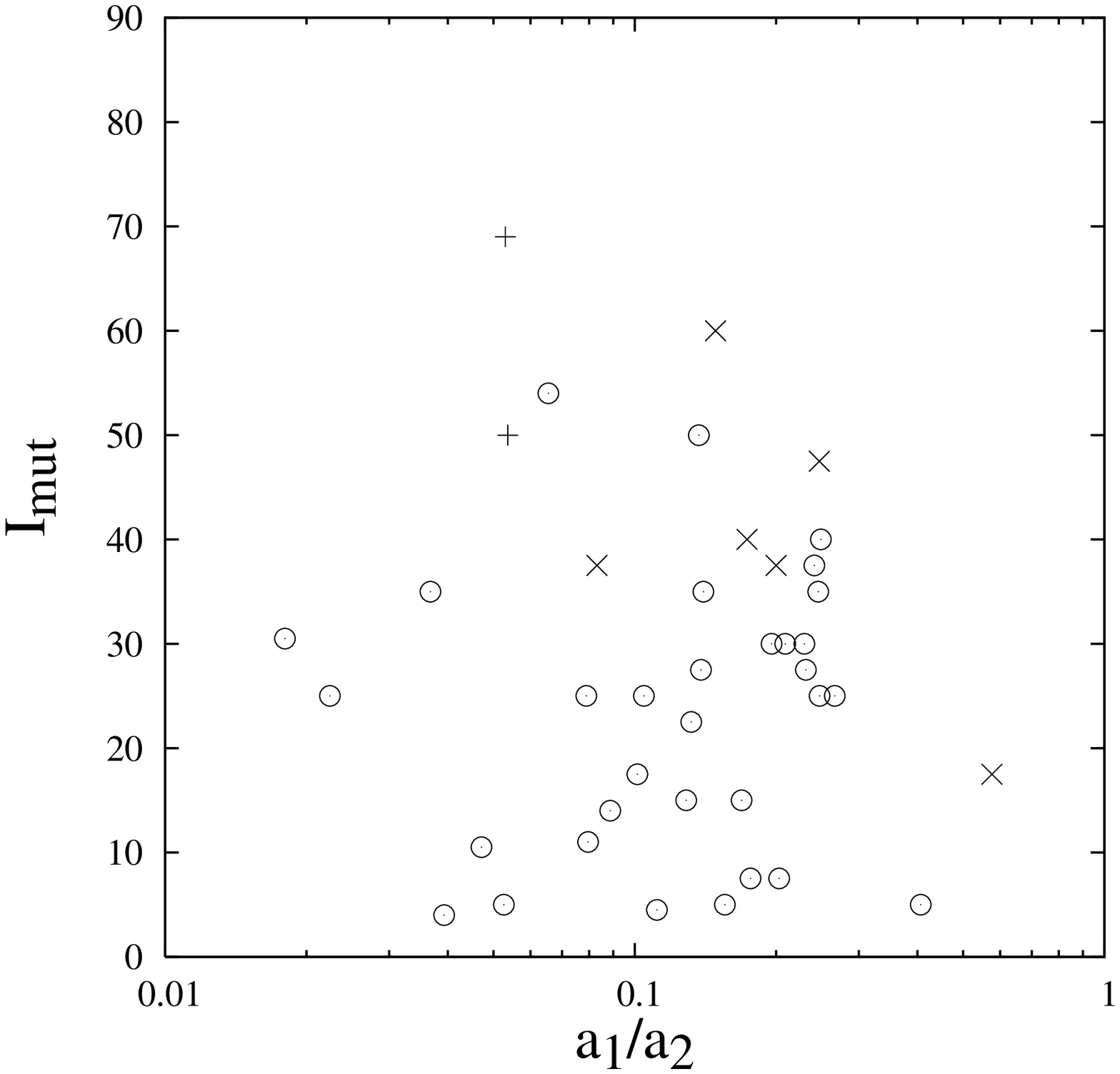}}
\caption{Top panel: Semi-major axis (in logarithmic scale) - eccentricity distribution of the surviving two-planet systems ('square' symbols for inner planets, 'circle' for outer planets). This distribution seems in agreement with the extrasolar planets  discovered so far. Bottom panel: Semi-major axes ratio (in logarithmic scale) - mutual inclination distribution of the surviving two-planet systems. These systems typically have large values of semi-major axes ratio ($\alpha = a_1/a_2 < 0.3$), while the mutual inclination can be as high as 70$^{\circ}$, with a median of $\sim 30^{\circ}$. The 'plus' symbol denotes stable Kozai-resonant systems, 'cross' unstable systems and 'circle' stable systems on the integration time.}
\label{twopl}
\end{figure}

The distribution of semi-major axis and eccentricity of the surviving planets can be seen in Fig. \ref{twopl}. As expected, this distribution is in agreement with previous results on planet-planet scattering (see e.g. \citealt{Cha07}), resembling the distribution of observed planets (although our runs are too few for a proper statistical comparison to be made). The distribution of the systems in terms of semi-major axis ratio, $\alpha$, and mutual inclination, $I_{\rm mut}$, is given in Fig.\ref{twopl}. As shown in this figure, $95\%$ of the systems have $\alpha < 0.3$. $20\%$ of the 40 systems are located outside the 7:1, 8:1 or even 9:1 MMR. The dynamical behavior of these systems is similar to that of the three first systems given in Table \ref{tableparam}, i.e.\ $\upsilon$ Andromedae, HD 12661 and HD 169830, which are outside the 5:1, 6:1 and 9:1 MMR, respectively (we refer to \citealt{Lib07a} for an analytical verification of the proximity of these systems to MMRs). Two systems are found close to the 5:2 and 4:1 MMR respectively, but no actual {\it resonant} system is found. This is in agreement with the work of  \citet{Cha07}, where only $1\%$ of the simulations resulted in temporary resonant capture (no dissipation was included in those simulations). Only two systems with high $\alpha$ -- similar to the  HD 155358 system of Table \ref{tableparam} -- were found. All the remaining systems are practically hierarchical systems, like the HD 74156 system in Table \ref{tableparam}.

\begin{figure}
\centering{
\hspace{-1cm}\rotatebox{270}{\includegraphics[height=15.cm]{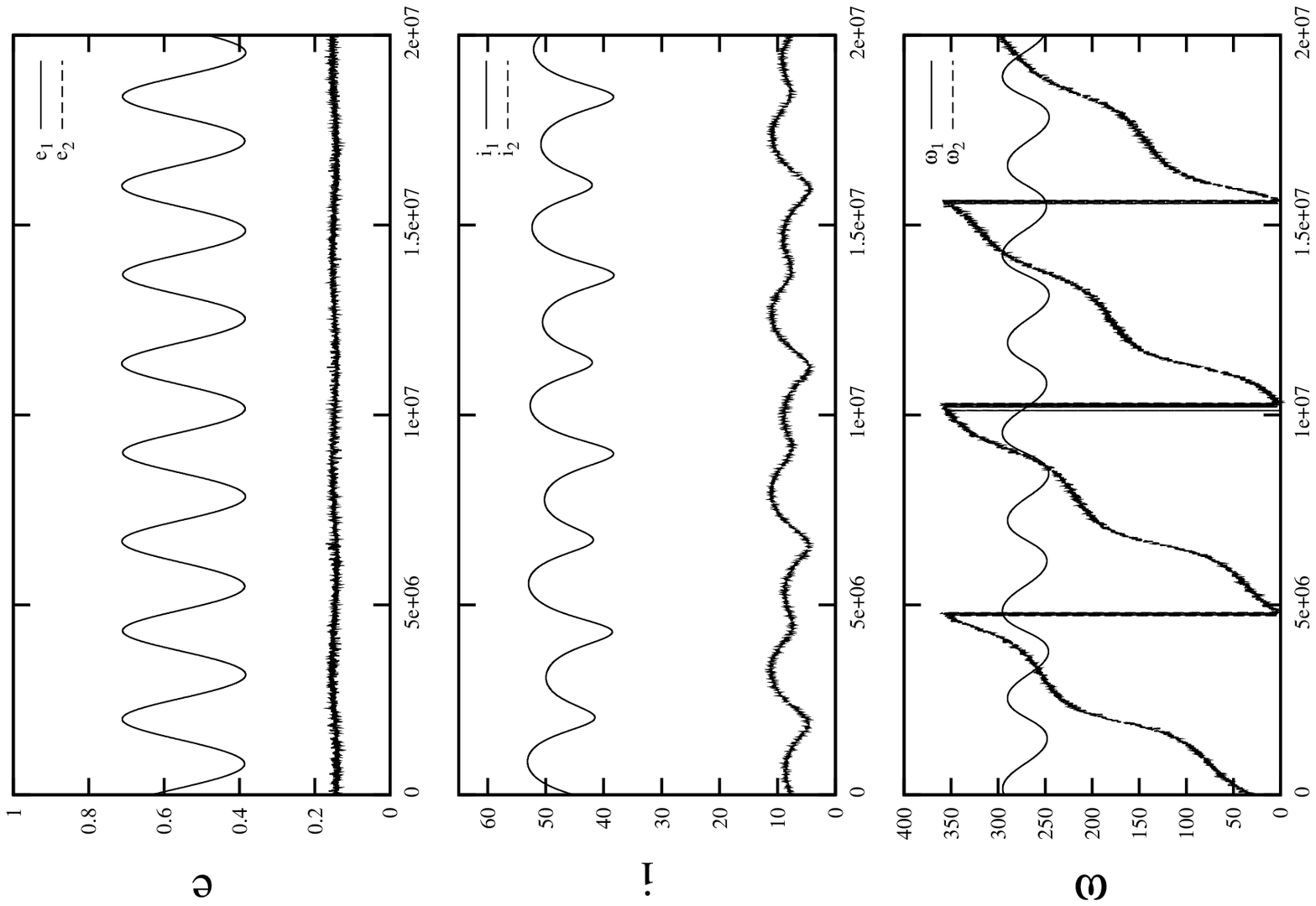}}
}
\caption{Example of a surviving two-planet system in a Kozai-resonant state whose dynamics is characterised by a coupled variation of the eccentricity and the inclination of the inner planet and a libration of the argument of the pericenter of the same planet around $270^\circ$.}
\label{kozai_fig}
\end{figure}

Of particular interest are 13 of the 40 surviving two-planet systems, which are highly non-coplanar ($I_{\rm mut}> 40^\circ$). These systems are in principle compatible with a Kozai resonance. Note that all these high-$I_{\rm mut}$ systems were obtained when $m_3$ was smaller than $m_1$ and $m_2$ (see Table \ref{tablescenario}). For these 13 systems, we performed an additional numerical integration for $2 \times 10^7~T_1$, in order to study their long-term stability. We found that 2 of these systems are actually in a stable Kozai-resonant state (see Fig. \ref{kozai_fig}), characterized by a coupled variation of eccentricity and inclination and libration of the argument of pericenter of the inner planet. Of the remaining systems, 6 have a stable secular behavior (not in Kozai resonance), while 5 systems became unstable (see Fig. \ref{twopl}). This instability is not related to close encounters but rather to the fact that the systems were located close to the separatrix that emanates from the unstable equilibrium of the Kozai resonance, encircling the libration zone (see \citealt{Mic06,Lib07b,Lib09a}). Let us note that Kozai resonance is not the only possible reason of secular instability, but in the present case, we have checked the five evolutions carefully and observed circulation of the argument of the pericenter alternately with libration around $90^\circ$ or $270^\circ$, showing the influence of the Kozai separatrix.

\section{Conclusions}

In this paper we examine a formation mechanism for ``3-D" planetary systems, composed of two (or more) giant planets. The starting point of this work is the assumption that systems with three or more planets can be driven by Type II migration into a multiply-resonant configuration, an evolution similar to the one recently proposed for our solar system. This delicate dynamical configuration can then become unstable, not because the planet are ``formed'' too close to each other (as is typically assumed in simulations of planet-planet scattering) but because the resonance can increase the eccentricities of the planetary orbits, up to the point that the orbits begin to intersect. 

We studied a set of nine different mass configurations. We find that a three-planet resonance (only the 1:2:4 relation is examined here) can be established, as long as the outer planet is not very massive, in which case the system dissolves before the multiple resonance is reached. In the opposite case (e.g. $m_3=m_2/2=m_1/2$), the resonance can even remain stable for a long time. Thus, as long as the disc dissipates quickly enough or exerts enough eccentricity damping on the planets as to prevent rapid eccentricity growth, a ``3-D" multi-resonant system can form. The inclinations of all planets can increase to values $>20^{\circ}$, if one of the eccentricities of the planets becomes $e\sim 0.5-0.6$. As found in our runs, the inclinations actually start growing only when the eccentricities of the planets become larger than $\sim 0.2$. This is reminiscent of the \citet{Tho03} mechanism, for two planets trapped in a 1:2 MMR. Our results suggest that this mechanism can be also active on three-planet configurations. 

In $\sim 90\%$ of our runs, the triple resonance is dynamically dissolved, as the planetary eccentricities grow because of the resonance, the orbits begin to intersect and the planets start encountering each other. Form then on, the evolution is typical of what is observed in planet-planet scattering simulations. About $50\%$ of the runs (in total) give rise to two-planet systems, while $\sim 25\%$ result in single-planet systems. An interesting feature of single-planet systems is that the final orbital inclination of the planet can be as large as $\sim 40^{\circ}$, with respect to the initial invariant plane (the disc plane). 

Two-planet systems, produced by this mechanism, are characterised by a large orbital separation of the two planets ($\alpha<0.3$ in $95\%$ of the cases); one planet is ejected on a hyperbolic orbit while the other two increase their orbital separation (the less massive planet moves inwards or outwards). The outer edge of the distribution ($\alpha_{\max}\approx 0.3$) seems relatively sharp, although this may be related to poor statistics. If not, it is interesting to check in the future whether this outcome is related to the resonance studied (i.e. the 1:2:4 resonance) and whether other resonant configurations produce different values of $\alpha_{\max}$. About $20\%$ of the systems have $\alpha \approx 0.3$, which is near the value of planets $c$ and $d$ of the $\upsilon-$And system. The distribution of mutual inclinations is broad, covering the range $0-70^{\circ}$. The median is $\sim 30^{\circ}$, which is also the value reported for the $\upsilon-$And system (\citealt{Mc10}). About $30\%$ of the systems have $I_{\rm mut}>40^{\circ}$. Finally, $5\%$ of the systems end up inside the libration zone of the Kozai resonance, while several more systems with  $I_{\rm mut}>40^{\circ}$ become unstable not because of planet-planet scattering but because they are near (yet, outside) the separatrix that encircles the libration zone of the Kozai resonance.

We conclude that the mechanism proposed here is quite robust in producing non-resonant ``3-D" systems, characterised by large mutual inclination (median $I_{mut}=30^{\circ}$) and large orbital separation ($95\%$ with $\alpha <0.3$).  Hydrodynamical simulations are needed in order to understand the conditions under which a system of three-planets can be trapped in a triple resonance -- and {\it which} resonant configurations are possible -- for different mass ratios. Although we believe that our work captures the essential qualitative dynamics of the studied mechanism, more work is needed (in the sense of more detailed modeling and a larger sample of runs) before accurate quantitative results are obtained. We reserve this work for a future publication.

\section{Ackowlegments}
The work of A-SL is supported by an FNRS post-doctoral research fellowship.

\bibliographystyle{}

\end{document}